\documentclass{elsart}

\usepackage{cite}
\usepackage{graphicx}

\begin{document}

\begin{frontmatter}

\title{On the application of the variational iteration method to a prey
and predator model with variable coefficients}

\author{Francisco M. Fern\'{a}ndez \thanksref{FMF}}

\address{INIFTA (UNLP,CCT La Plata-CONICET), Divisi\'{o}n Qu\'{i}mica Te\'{o}rica,\\
Diag. 113 y 64 (S/N), Sucursal 4, Casilla de Correo 16,\\
1900 La Plata, Argentina}

\thanks[FMF]{e--mail: fernande@quimica.unlp.edu.ar}

\begin{abstract}
We discuss an amazing prey--predator model with variable
coefficients, analyze its predictions and the accuracy of the variational
iteration method used to solve the nonlinear equations.
\end{abstract}

\end{frontmatter}

\section{Introduction}

There has recently been great interest in the application of several
approximate procedures, like the homotopy perturbation method (HPM), the
Adomian decomposition method (ADM), and the variation iteration method
(VIM), to a variety of linear and nonlinear problems of interest in
theoretical physics \cite
{RDGP07,CHA07,YO07,CH07a,EG07,GAHT07,SNH07,CH08,ZLL08,SNH08b,M08,SNH08,RAH08,YE08,SG08}%
. In a series of papers I have shown that most of the results produced by
those methods are useless, nonsensical, and worthless\cite{F07,F08b,F08c}.
From now on I will refer to those variation and perturbation approaches as
VAPA.

The purpose of this communication is to analyze a recent application of the
VIM to a model for the evolution of a prey--predator system\cite{YE08} .

\section{The model}

Yusufo\u{g}lu and Erba\c{s}\cite{YE08} have recently applied the VIM to a
prey--predator model with variable coefficients
\begin{eqnarray}
\dot{x}(t) &=&a(t)x(t)-b(t)x(t)y(t),\;x(0)=\alpha  \nonumber \\
\dot{y}(t) &=&-c(t)y(t)+d(t)x(t)y(t),\;y(0)=\beta  \label{eq:model}
\end{eqnarray}
According to the authors $x(t)$ and $y(t)$ represent the populations of
rabbits and foxes, respectively. Instead of constants $a,b,c,d$ for the
growth rate of the prey, the efficiency of the predator's ability to capture
the prey, the death rate of the predator, and the growth rate of the
predator, the authors introduce functions of time\cite{YE08} . The authors
do not mention any physical, zoological or ecological reason for the
substitution of functions for constants, and at first sight it seems
arbitrary and unjustified.

In the first example Yusufo\u{g}lu and Erba\c{s}\cite{YE08} choose $%
a(t)=b(t)=-t$, $c(t)=d(t)=t$, $\alpha =\beta =2$. In principle, there is no
reason for this choice of time--dependent coefficients, except that the
equations can be solved exactly. The exact solutions for this model
\begin{eqnarray*}
x(t) &=&\frac{2}{[2-\exp (t^{2}/2)]} \\
y(t) &=&\frac{2}{[2-\exp (t^{2}/2)]}
\end{eqnarray*}
predict that the population of rabbits always equals the foxes one and both
tend to infinity as $t$ approaches $t_{c}=\sqrt{2\ln 2}$ from the left. The
reader may think that this behaviour of the model is completely unrealistic
because no population becomes infinity in the world we perceive. However,
Yusufo\u{g}lu and Erba\c{s}\cite{YE08} did not find it unreasonable, and for
the time being we may assume that the planet is infinitely large and can
accommodate infinite populations of rabbits and foxes. Or that it takes
place in another world where such a curious behavior may be possible. It is
more difficult to understand that when $t>t_{c}$ both populations jump to $%
-\infty $ and then approach zero from below as $t\rightarrow \infty $. I am
not smart enough to explain this behaviour of the extraterrestrial rabbits
and foxes and therefore I have decided to describe it without further
analysis.

Yusufo\u{g}lu and Erba\c{s}\cite{YE08} applied the VIM and obtained
increasingly accurate approximations $x_{j}(t)$ and $y_{j}(t)$ for $j=1,2,3$%
; for example
\begin{eqnarray}
x_{3}(t) &=&256+\left( \frac{1882}{3}-160t^{2}-64t^{4}\right)
e^{-t^{2}/2}-\left( 72+32t^{2}\right) e^{-3t^{2}/2}  \nonumber \\
&&-\left( 804+416t^{2}+64t^{4}\right) e^{-t^{2}}-\frac{16}{3}e^{-2t^{2}}
\label{eq:ex1_x3}
\end{eqnarray}
We appreciate that this approximate solution does not exhibit the pole of
the exact answer, and tends to zero as $t\rightarrow \infty $. Yusufo\u{g}lu
and Erba\c{s}\cite{YE08} cleverly overcome this difficulty by comparing
their approximate expression and the exact solution only within the interval
of reasonable agreement. Besides, notice that the VIM corrects the curious
behaviour of the extraterrestrial rabbits and foxes and makes them extinct
after a sufficiently long time.

Naively, I tried Pad\'{e} approximants and found that the simplest ones give
much better results than the elaborated combinations of polynomials and
exponentials of Yusufo\u{g}lu and Erba\c{s}\cite{YE08}. The fact is that the
$t$--power series expansion of the solution converges for all $t<t_{c}$, and
the Pad\'{e} approximants constructed from it take into account the poles of
the solution as zeros of the denominator~\cite{BO78}. For example, the
simple and straightforward $[2/4]$ Pad\'{e} approximant
\begin{equation}
\lbrack 2/4](t)=\frac{8(t^{2}-6)}{(t^{4}+16t^{2}-24)}
\label{eq:example1[3/4]}
\end{equation}
built from the time--series solutions to the model equations yields much
more accurate results than the VIM equation (\ref{eq:ex1_x3}). It seems to
be considerably easier to obtain the power series and their Pad\'{e}
approximants than the application of the VIM. However, you cannot publish
the much more reasonable former approach because it is just a textbook
example for students. It is worth mentioning that the approximants $[1/2](z)$%
, $[2/3](z)$, $[3/4](z)$ and $[4/5](z)$, $z=t^{2}$, exhibit poles at $%
t^{2}=1.380831519$, $t^{2}=1.386322332$, $t^{2}=1.38629429$, and $%
t^{2}=1.386294361$, respectively, that clearly converge towards the exact
pole $t_{c}^{2}=1.386294364$. The VIM solutions do not exhibit this property
and therefore fail to follow the exact solution as $t$ increases. However, I
have been much surprised at the revelation that part of the scientific
community does not want to be bothered by such mathematical details.

The second example clearly shows that the unparalleled insight of Yusufo\u{g}%
lu and Erba\c{s}\cite{YE08} is beyond any mortal's perception of the very
nature of the universe. They masterfully choose
\begin{eqnarray}
a(t) &=&4+\tan t,b(t)=\exp (2t),c(t)=-2,d(t)=\cos t,  \nonumber \\
\alpha  &=&-4,\;\beta =4  \label{eq:example2}
\end{eqnarray}
First of all, notice the initial negative population of rabbits $\alpha =-4$%
!!  The exact solutions
\begin{eqnarray}
x(t) &=&-4/\cos t,  \nonumber \\
y(t) &=&4e^{-2t}  \label{eq:example2_x_y}
\end{eqnarray}
show that the population of rabbits remains negative and tends to $-\infty $
as $t\rightarrow (\pi /2)^{-}$, then it jumps to plus infinity and starts
decreasing as $t$ increases. On the other hand, the population of foxes
tends to zero exponentially, probably due to the stress caused by the
negative population of rabbits and their sudden emerging to real world.

Once again, Yusufo\u{g}lu and Erba\c{s}\cite{YE08} apply VIM and obtain
third--order approximate populations. I beg the reader to have a look at the
authors' Fig. 2 to appreciate the remarkable performance of the VIM. If that
science overload does not exhaust the reader's mind, he/she may then compare
the approximate expressions of Yusufo\u{g}lu and Erba\c{s}\cite{YE08} with
the naive Pad\'{e} approximants

\begin{eqnarray}
x[2/4](t) &=&\frac{-16(t^{2}+30)}{3t^{4}-56t^{2}+120}  \nonumber \\
y[3/4](t) &=&\frac{-4(4t^{3}-30t^{2}+90t-105)}{%
2t^{4}+16t^{3}+60t^{2}+120t+105}  \label{eq:example2[3/4]}
\end{eqnarray}
which are far simpler than the VIM solutions and give much better results.
For that reason they are unsuitable for publication.

Yusufo\u{g}lu and Erba\c{s}\cite{YE08} also applied the VIM to models with
constant coefficients. Those models are more realistic from a zoological
point of view. However, the authors do not show their results and simply
mention that they agree with those obtained by the HPM\cite{RDGP07}. For
example, case I is given by $a=b=1$, $d=10c=1$, $\alpha =14$, $\beta =18$%
\cite{YE08,RDGP07}. This model exhibits two stationary points: a saddle
point at $(x_{s},y_{s})=(0,0)$ and a center at $(x_{s},y_{s})=(1/10,1)$. The
authors do not attempt to reproduce the overall picture of the model
dynamics, which is what really matters\cite{BO78}, and restrict themselves
to a time interval about the origin because the VIM results will prove
entirely useless otherwise.

In Fig. 1 we show the populations given by the expressions derived by Rafei
et al\cite{RDGP07} by means of the HPM and the exact (numerical) results. We
clearly appreciate that both the VIM (Fig. 3 of Yusufo\u{g}lu and Erba\c{s}%
\cite{YE08}) and the HPM\cite{RDGP07} are far from giving a reasonable
picture of the behaviour of the prey and predator populations. The same
conclusion holds for the other cases treated by Rafei et al\cite{RDGP07}.
However, if you restrict to the initial time when the animals begin to make
acquaintances, then the VAPA results\cite{CHA07,RDGP07,YE08} are not too bad
(see my earlier discussion of the subject\cite{F07}). The fact that nobody
in the field of population dynamics is interested in the initial evolution
of the system has remained unnoticed in most VAPA applications\cite
{CHA07,RDGP07,YE08}.

\section{Conclusions}

It is amazing the amount of nonsensical VAPA\ papers that have recently been
published on the treatment of all kinds of linear and nonlinear problems. It
is surprising the increasing interest of part of the scientific community in
remarkably useless results. The list below shows only those I had time to
peruse. The work discussed in this communication is just an example. The
reader may also have a look at my previous analysis of other papers\ref
{F07,F08b,F08c}.

\begin{figure}[H]
\begin{center}
\includegraphics[width=9cm]{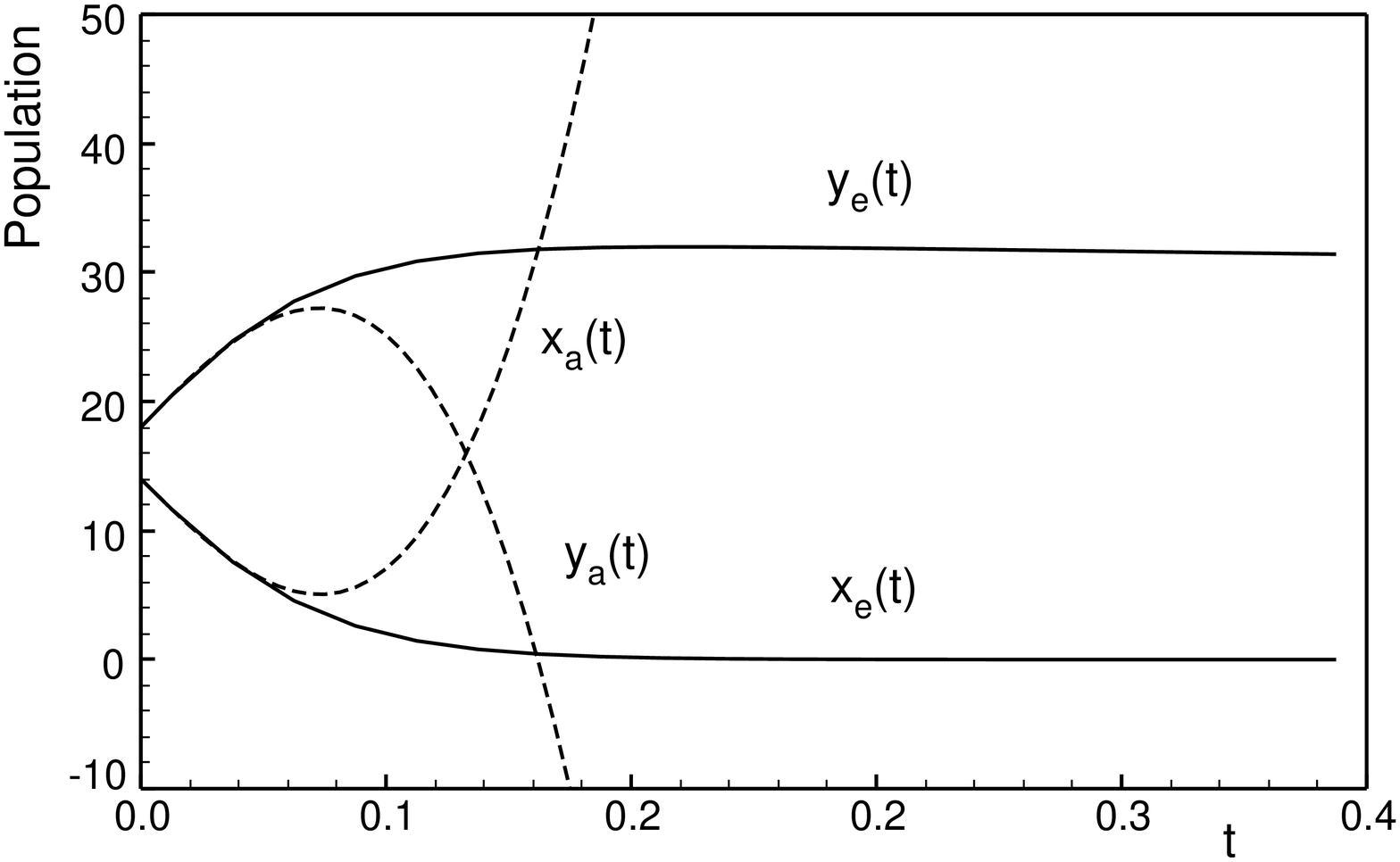}
\end{center}
\caption{Exact (e) and approximate (a) populations for the model with
constant coefficients}
\label{Fig:VIMC1}
\end{figure}

\end{document}